\documentclass[twocolumn]{aastex63}

\newcommand{\obj}{NGC 4395}
\def\CIV{C\,{\sc iv}}

\submitjournal{ApJ}

\shorttitle{Optical Variability of NGC 4395}
\shortauthors{Burke et al.}
\graphicspath{{./}{figures/}}

\begin{document}

\title{Optical Variability of the Dwarf AGN NGC 4395 from the Transiting Exoplanet Survey Satellite}

\correspondingauthor{Colin J. Burke, Yue Shen}
\email{colinjb2@illinois.edu, shenyue@illinois.edu}

\author[0000-0001-9947-6911]{Colin J. Burke}
\affiliation{Department of Astronomy, University of Illinois at Urbana-Champaign, 1002 W. Green Street, Urbana, IL 61801, USA}
\affiliation{National Center for Supercomputing Applications, 
1205 West Clark Street, Urbana, IL 61801, USA}

\author[0000-0003-1659-7035]{Yue Shen}
\altaffiliation{Alfred P. Sloan Fellow}
\affiliation{Department of Astronomy, University of Illinois at Urbana-Champaign, 1002 W. Green Street, Urbana, IL 61801, USA}
\affiliation{National Center for Supercomputing Applications, 
1205 West Clark Street, Urbana, IL 61801, USA}

\author[0000-0002-9932-1298]{Yu-Ching Chen}
\affiliation{Department of Astronomy, University of Illinois at Urbana-Champaign, 1002 W. Green Street, Urbana, IL 61801, USA}

\author[0000-0001-5387-7189]{Simone Scaringi}
\affiliation{Department of Physics and Astronomy, Texas Tech University, PO Box 41051, Lubbock, TX 79409, USA}

\author[0000-0002-4900-6628]{Claude-Andre Faucher-Giguere}
\affiliation{Department of Physics and Astronomy and CIERA, Northwestern University, 2145 Sheridan Road, Evanston, IL 60208, USA}

\author[0000-0003-0049-5210]{Xin Liu}
\affiliation{Department of Astronomy, University of Illinois at Urbana-Champaign, 1002 W. Green Street, Urbana, IL 61801, USA}
\affiliation{National Center for Supercomputing Applications, 
1205 West Clark Street, Urbana, IL 61801, USA}


\author[0000-0002-6893-3742]{Qian Yang}
\affiliation{Department of Astronomy, University of Illinois at Urbana-Champaign, 1002 W. Green Street, Urbana, IL 61801, USA}




\begin{abstract}

We present optical light curves from the Transiting Exoplanet Survey Satellite (TESS) for the archetypical dwarf active galactic nucleus (AGN) in the nearby galaxy NGC 4395 hosting a $\sim 10^5\,M_\odot$ supermassive black hole (SMBH). Significant variability is detected on timescales from weeks to hours before reaching the background noise level. The $\sim$month-long, 30 minute-cadence, high-precision TESS light curve can be well fit by a simple damped random walk (DRW) model, with the damping timescale $\tau_{\rm DRW}$ constrained to be $2.3_{-0.7}^{+1.8}$~days ($1\sigma$). NGC 4395 lies almost exactly on the extrapolation of the $\tau_{\rm DRW}-M_{\rm BH}$ relation measured for AGNs with BH masses that are more than three orders of magnitude larger. The optical variability periodogram can be well fit by a broken power law with the high-frequency slope ($-1.88\pm0.15$) and the characteristic timescale ($\tau_{\rm br}\equiv 1/(2\pi f_{\rm br})=1.4_{-0.5}^{+1.9}\,$days) consistent with the DRW model within 1$\sigma$. This work demonstrates the power of TESS light curves in identifying low-mass accreting SMBHs with optical variability, and a potential global $\tau_{\rm DRW}-M_{\rm BH}$ relation that can be used to estimate SMBH masses with optical variability measurements. 



\end{abstract}



\section{Introduction} \label{sec:intro}

\obj\ is a well-known nearby dwarf galaxy (at a distance of $\sim 4$\,Mpc) that hosts a Seyfert 1 nucleus with its black hole (BH) mass estimated in the range of $10^4-10^6\,M_\odot$ \citep[][]{Filippenko_Ho_2003,Peterson_etal_2005,Vaughan_etal_2005,denBrok_2015,Brum_2019,Woo_etal_2019} using various techniques. Optical variability from the nucleus of \obj\ has been clearly detected on short timescales (less than a day), which facilitated several reverberation mapping campaigns to measure its BH mass \citep[e.g.,][]{Peterson_etal_2005,Woo_etal_2019}. Given its small distance, \obj\ is bright enough ($i\approx14$) for the Transiting Exoplanet Survey Satellite (TESS) to detect the variability from the unobscured nucleus.

The significance of studying \obj\ with TESS light curves is two-fold. First, this system is the best example to test the capability of the high-cadence, precision TESS light curves to discover tenuous nuclear variability due to the presence of a low-luminosity active galactic nucleus (AGN), in particular for bright nearby galaxies among which many are in the dwarf galaxy regime. This variability technique has proven successful in identifying accreting supermassive black holes (SMBHs) both for distant, high-luminosity quasars \citep[e.g.,][]{Butler_Bloom_2011}, and for nearby, low-luminosity AGNs, and hence low-mass, accreting SMBHs \citep[e.g.,][]{Baldassare_etal_2018}. Discovering low-mass SMBHs (e.g., $\lesssim 10^6\,M_\odot$) and furthermore, measuring their occupation fraction among low-mass galaxies, are of critical importance to understanding the seeding scenarios of SMBHs \citep[e.g.,][]{Greene_etal_2019} and the co-evolution between central BHs and host galaxies \citep[e.g.,][]{Kormendy_Ho_2013}. Motivated by the increasing availability of light curves from optical surveys, several recent studies have identified dwarf AGNs using optical variability \citep[e.g.,][]{Baldassare_etal_2018,Martinez-Palomera_2020,Secrest_2020}. However, due in part to the limited cadence of these surveys and the signal-to-noise ratio of the intrinsic variability, the selection rates are generally low (a few percent) and the sampling is insufficient to study the variability properties of dwarf AGNs at days to hours timescales. Being one of the smallest SMBHs with well-measured masses, \obj\ is of particular significance and is the poster-child for testing variability identification of dwarf AGNs with high-cadence data.

Secondly, the high-quality (duration, cadence, precision) TESS light curves will enable one of the best measurements of the optical variability characteristics of \obj. The advent of time-domain optical imaging surveys in recent years has greatly improved the study of AGN variability. The AGN optical emission mostly probes the accretion disk in efficiently accreting SMBHs, extending from $\sim$a few gravitational radii ($R_{g}\equiv GM/c^2$) to perhaps thousands of $R_g$. The detailed variability properties as measured by structure functions, power spectral density (PSD), variable flux distributions, etc., can be used to constrain models of the accretion disk. Unlike X-ray variability studies of AGNs, high-quality measurements of detailed optical light curves and PSDs of AGNs are just beginning \citep[e.g.,][]{MacLeod_etal_2010,Mushotzky_etal_2011,Simm_etal_2016,Caplar_etal_2017,Smith_etal_2018}. 

One of the most important discoveries from these optical variability studies is that the stochastic AGN optical light curves can be well described by a damped random walk (DRW) model that features a transition from a random walk PSD ($\propto f^{-2}$) on short timescales to a white noise PSD on long timescales \citep[e.g.,][]{Kelly_etal_2009,Kozlowski_etal_2010,MacLeod_etal_2010,MacLeod_etal_2012}. Deviations from this simple DRW model have been reported, mostly on the shortest timescales where the PSD seems to steepen, indicative of reduced variability \citep[e.g.,][]{Mushotzky_etal_2011,Kasliwal_etal_2015}. 

While the DRW model is only an empirical model to describe the optical variability of AGNs, it is tempting to relate its model parameters to physical properties of the accretion disk. For example, \citet{Kelly_etal_2009} demonstrated that the damping timescale $\tau_{\rm DRW}$, corresponding to the transition frequency ($f_{\rm DRW}=(2\pi\tau_{\rm DRW})^{-1}$) between the red noise and white noise in the PSD, scales with the BH mass for a sample of AGNs. This is expected if this damping timescale parameter is related to physical variability processes in the accretion disk. Since most timescales in the accretion disk scale with the mass of the BH, a trend with BH mass is a natural consequence. Given the extremely low BH mass of \obj, it is then interesting to see if it has a much shorter damping timescale than those of its high-mass counterparts. Using the empirical relation measured in \citet{Kelly_etal_2009}, the expected damping timescale decreases from $\sim 200$ days for luminous quasars ($M_{\rm BH}\sim 10^{8}\,M_\odot$) to a few days for $M_{\rm BH}\sim 10^5\,M_\odot$, a regime that is well sampled by the TESS light curves from a single sector.


In this work we report our time series analysis of the TESS light curves for \obj, as part of a much larger effort to use TESS data to study nuclear variability of galaxies near and far (Y.~Shen et~al., in preparation). We describe the technical details of producing TESS light curves for \obj\ in \S\ref{sec:methods} (with additional information presented in the appendices), present our results in \S\ref{sec:results} and conclude in \S\ref{sec:discussion}. All uncertainties are $1\sigma$ unless otherwise specified, and all logarithms are 10 based.

\section{Data} \label{sec:methods}


TESS \citep{Ricker_etal_2014} is a NASA Explorer-class mission designed to image nearly the entire sky to search for exoplanets, particularly around M dwarf stars, using the transit method. The spacecraft has four wide-field ($24^\circ{\times}24^\circ$ unvingetted) optical charge-coupled device (CCD) cameras with an image scale of 21 arcsec pixel$^{-1}$ (angular resolution about 42\arcsec\ in FWHM) and a wide $600$ -- $1000$ nm bandpass. The standard TESS cadence is 30 minutes throughout a 30 day fixed-pointing ``sector''. The reduced full-frame images (FFIs) are publicly released for each sector\footnote{The data described here may be obtained from the MAST archive at \url{https://archive.stsci.edu/tess}.}. \obj\ was observed in TESS sector 22 with camera 2, CCD 4 during the month of 2020 March.

We have searched the literature for BH mass estimates of the AGN in \obj. Most previous estimates are a few times $10^5\,M_\odot$ based on UV (broad \CIV\ line) reverberation mapping \citep{Peterson_etal_2005}, X-ray variability PSD \citep[e.g.,][]{Shih_etal_2003,Vaughan_etal_2005}, stellar properties of the host galaxy \citep[e.g.,][]{Filippenko_Ho_2003}, and gas kinematics \citep{denBrok_2015,Brum_2019}. A more recent reverberation mapping study using the optical broad lines reported a BH mass of $\sim 10^4\,M_\odot$ \citep{Woo_etal_2019}. We adopt a fiducial BH mass of $10^5\,M_\odot$ throughout this work as an approximate average of these measurements. Despite being one of the least luminous AGNs, with an estimated Eddington ratio of $\sim 10^{-3}$ \citep[e.g.,][]{Peterson_etal_2005}, \obj\ is nevertheless a classic Seyfert 1 AGN with notable optical variability down to intra-day timescales. 







\begin{figure*}
\gridline{\fig{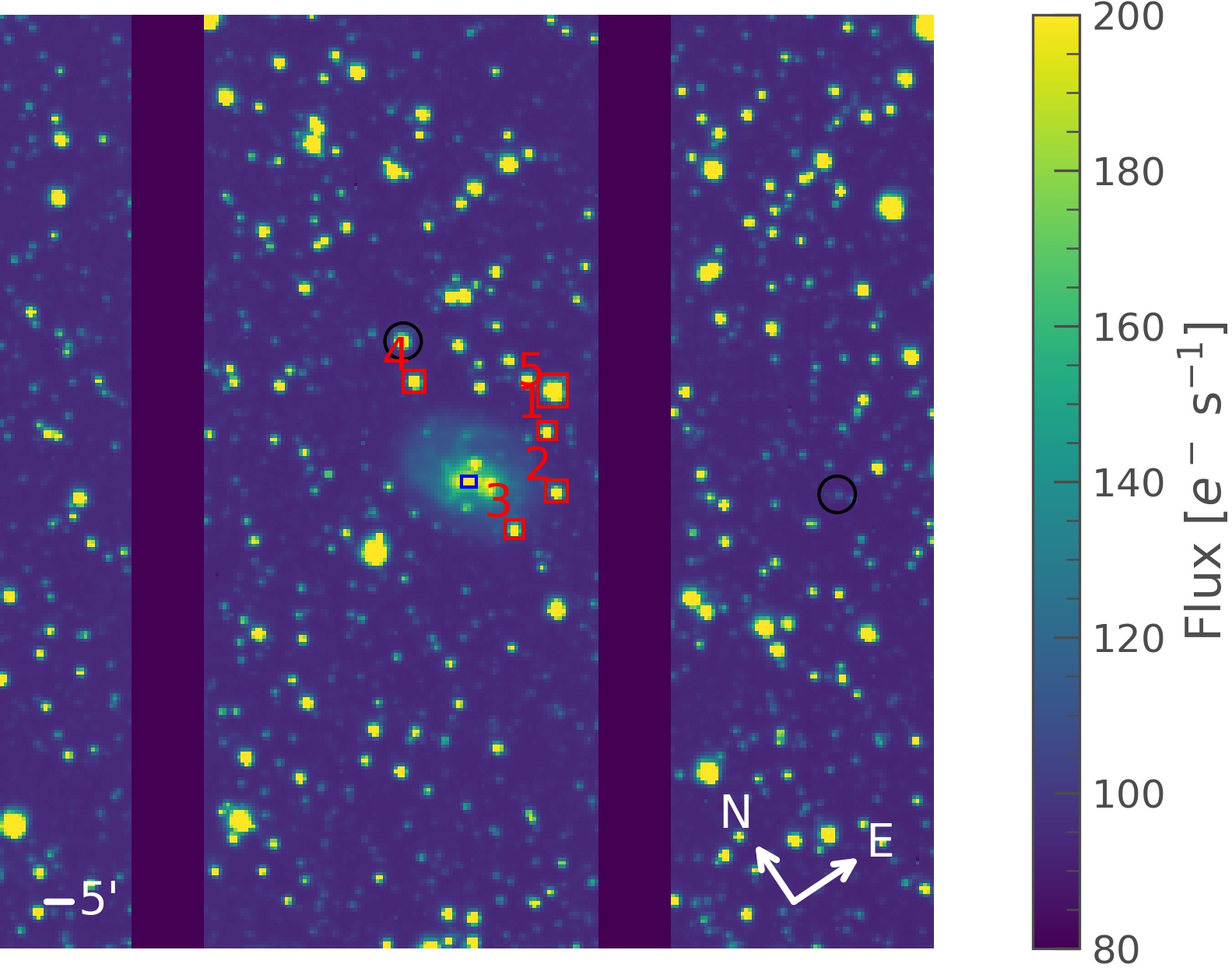}{0.485\textwidth}{(a)}
          \fig{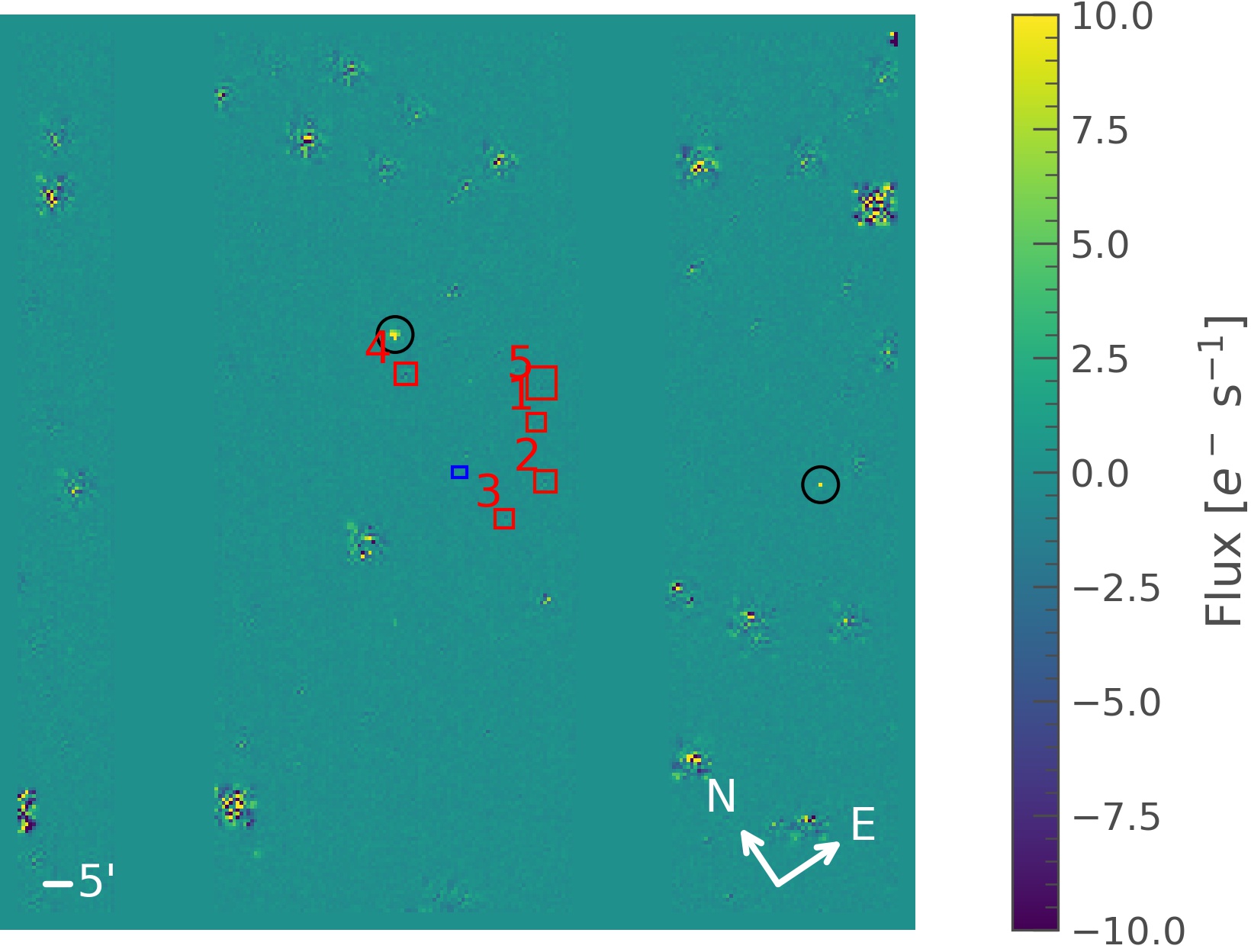}{0.5\textwidth}{(b)}
          }
\caption{Template (a) and example difference (b) frames ($256\times256$ pixels) of sector 22. The target aperture (blue) is centered on the nucleus of NGC 4395 (blue). Nearby comparison star apertures are shown in red. Columns affected by noise from charge-transfer straps have been masked out (i.e., the two dark vertical stripes). Artifact flux residuals can be seen near several bright sources. Real flux residuals can also be seen for two variable stars in this field (black circles). \label{fig:images}}
\end{figure*}

\begin{figure*}
\fig{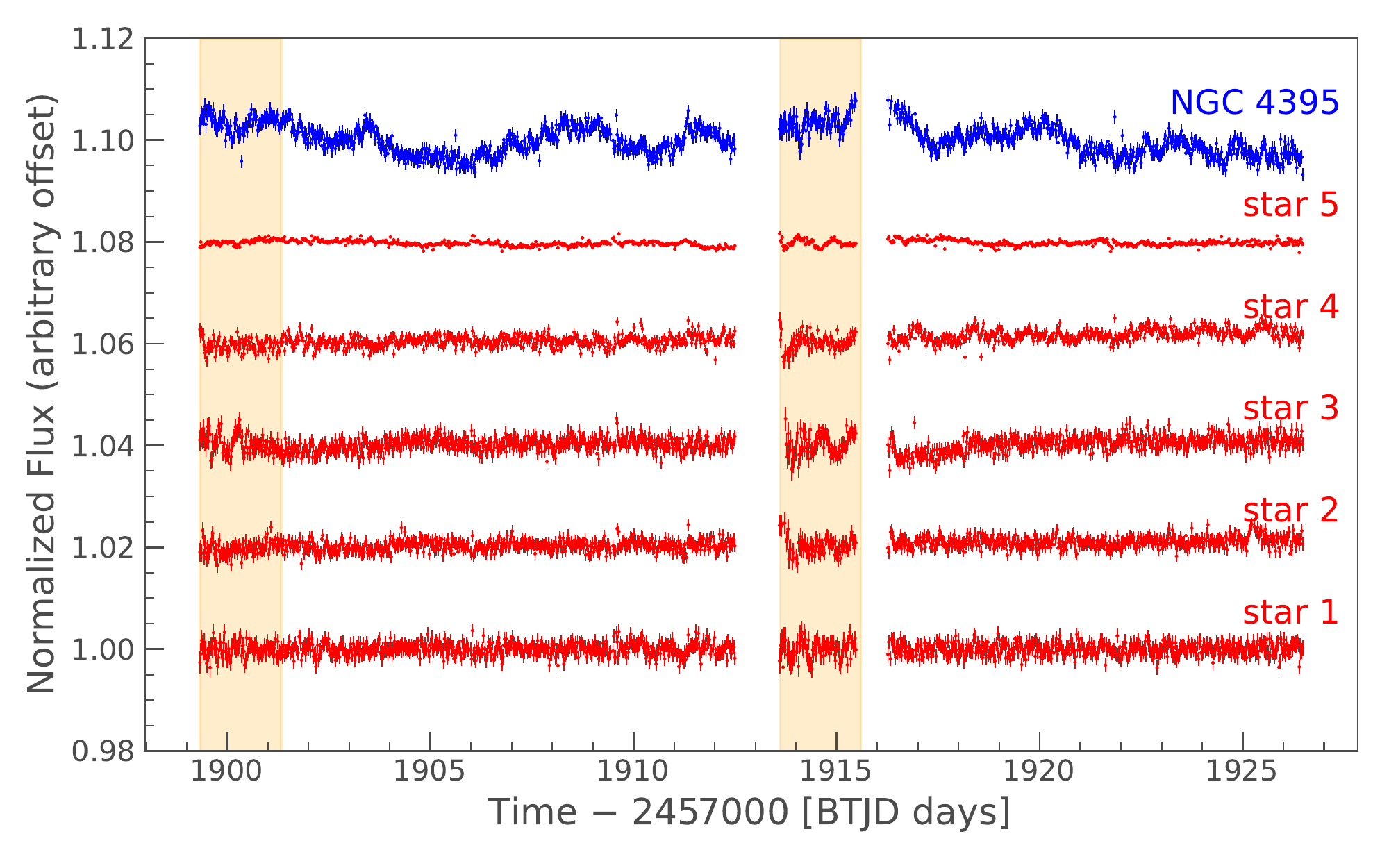}{0.95\textwidth}{}
\caption{Final TESS DIA light curves of the target (blue, top row) and nearby field stars (red) shown on a normalized flux scale. The orange shaded areas show the first day of each orbit: the thermal recovery period when residual systematic trends tend to be large. The empirical corrections applied to the light curves are described in Appendix~\ref{sec:detrending}. \label{fig:lc_corrected}}
\end{figure*}

\subsection{Light-curve Extraction}

Despite the overall superb photometric precision, TESS images suffer from many systematics (e.g., scattered light, spacecraft pointing jitter, rolling-band noise; see \citealt{Vanderspek_etal_2018}) which can obscure or even mimic weak AGN-like variability. Extreme care must be taken to correct these systematic trends without subtracting any significant real astrophysical signal. To mitigate TESS systematics, we employ difference imaging analysis (DIA) using FFIs, similar to the method described by \citet{Oelkers_Stassun_2018}. For extended sources like \obj\ and given the large TESS pixels, DIA is necessary to accurately remove the constant flux from the host galaxy and superposing foreground stars, and reveal the intrinsic variability from the nucleus. 

Our custom DIA analysis pipeline makes use of \textsc{hotpants} \citep{Becker_2015} based on the algorithm in \citet{Alard_Lupton_1998} and \citet{Alard_2000}, and is built on top of the \textsc{Lightkurve} code \citep{Lightkurve_Collaboratioin_2018}. Below we summarize the main steps of our pipeline:

\begin{enumerate}
    \item Query TESS FFIs given the target coordinates.
    \item Cutout the FFI around the target coordinates and reproject each FFI to match the reference frame world coordinate system (WCS). This corrects for some of the spacecraft pointing drift throughout a TESS observation provided it is captured in the WCS. We use \textsc{reproject}'s flux-conserving \texttt{reproject\_exact} algorithm. We found a cutout size of $256{\times}256$ pixels sufficient to find multiple bright reference stars.
    \item Mask out columns with enhanced noise due to charge-transfer straps behind the CCD. This time-varying systematic is caused by red scattered light from the Earth penetrating the CCD and reflecting off of charge-transfer straps behind the CCDs. Thankfully, \obj\ is not near any of the impacted columns, therefore we mask the affected pixels.
    \item Find field stars in the frame to use as \textsc{hotpants} sub-stamp positions. We make use of \textsc{photutils}' \texttt{DAOStarFinder} method \citep{Stetson_1987}. The sources inside each sub-stamp are used to build the point-spread function (PSF) kernel (see Appendix~\ref{sec:kernel}). We choose the sources by generating preliminary simple aperture photometry (SAP) light curves for each candidate and reject any with discernible intrinsic variability by eye.
    \item Make a template frame using a coadd of the FFI cutouts. We use a median-combined coadd of 20 frames centered on the first orbit, where the PSF and background are stable. This is consistent with the findings of \citet{Alard_Lupton_1998} that a template frame consisting of a coadd of the best 20 frames averages out variations in the PSF and results in reduced noise amplitude in DIA light curves.
    \item Difference each FFI cutout with the template frame after convolving the template to match the PSF size of the FFI cutout using \textsc{hotpants}. Background subtraction is also done at this stage using 3$\sigma$ sigma-clipped median calculated in a $\sim100\times100$ pixels stamp inside each image. The background is allowed to vary linearly across each stamp as a 2D gradient. The reference frame and an example of a difference frame are shown in Figure~\ref{fig:images}.
    \item Coadd the difference and template frames. We then convert the difference and coadded frames to new \textsc{Lightkurve} \texttt{TargetPixelFile} objects. These frames are convolved to match the PSF of the template and normalized to the flux level of the template. This step results in background-subtracted frames with a uniform PSF. The sub-stamp normalization procedure also acts as an initial detrending of the images.
\end{enumerate}

The final \texttt{TargetPixelFiles} are nearly science-ready, and we use field stars to confirm the variability of the target is real\footnote{We also generated an animated movie of the difference frames to confirm the quality of image subtraction. It can be viewed at \url{http://quasar.astro.illinois.edu/tess/ngc4395.gif}.}. Five field stars were chosen for comparison, as indicated in Figure~\ref{fig:images}. We selected these stars because they are relatively bright, isolated, and nearby ($\lesssim12^{\prime}$ away from the target).


The last step is detrending residual trends in the light curves due to instrument systematics. We use star 1 as the calibration star because it is impacted most strongly by the common systematic dips displayed in the light curves (see Appendix \ref{sec:detrending}). We fit a spline model to trends in its light curve. We then fit a re-scaled version of this model to the light curves of the target and other field stars. This method preserves the intrinsic variability of the target while correcting for common trends in the calibration field star light curve. The full method is described in Appendix~\ref{sec:detrending}. The final detrended light curves for \obj\ and comparison stars are shown in Figure~\ref{fig:lc_corrected}. We confirm that using the original, uncorrected light curves yields consistent results in our light-curve analysis below.



Given the brightness of \obj\ ($\approx 13-14$ mag), the expected background-limited photometric precision is about $\sim 0.2\%$, roughly consistent with the level of uncertainties in our DIA light curve.

The large pixel size (and PSF) of TESS forced us to use a large aperture ($4\times3$ pixels) to enclose most of the variable nuclear flux. We confirmed that using alternate apertures of $3\times3$ and $4\times4$ pixels produced consistent results. Inevitably a large fraction of the host galaxy is included in the TESS aperture. While the variable flux from DIA measures the nuclear-only variability, the baseline aperture flux measured from the reference frame includes the dilution from host light. This constant host contamination does not affect the shape of the PSD, but changes the fractional variability compared to other measurements using different apertures.


The extracted TESS light curves (in different forms, e.g., DIA, SAP, corrected/uncorrected for scattered light in the first day of each orbit) for \obj\ are presented in Table~\ref{tab:1}. The SAP light curve verifies the basic variability patterns observed in the DIA light curve, but has additional long-term systematics due to improper background subtraction for extended sources such as \obj. Therefore we use the corrected DIA light curves as the default for subsequent analysis. 

\begin{deluxetable*}{ccccccc}
\tablenum{1}
\tablecaption{Extracted TESS \obj\ Light Curves}
\tablewidth{0pt}
\tablehead{
\colhead{Time} & \colhead{Flux DIA} & \colhead{Flux Err. DIA} & \colhead{Flux DIA uncorr.} & \colhead{Flux Err. DIA uncorr.} & \colhead{Flux SAP} & \colhead{Flux Err. SAP} \\
\colhead{[BTJD $-$ 2457000]} & \colhead{[$e^-$ s$^{-1}$]} & \colhead{[$e^-$ s$^{-1}$]} & \colhead{[$e^-$ s$^{-1}$]} & \colhead{[$e^-$ s$^{-1}$]} & \colhead{[$e^-$ s$^{-1}$]} & \colhead{[$e^-$ s$^{-1}$]} 
}

\decimalcolnumbers
\startdata
 1899.327623 &  1590.755493 &      2.868915 &      1576.835693 &             2.868915 &  1626.485352 &      2.771091 \\
 1899.348457 &  1591.385742 &      2.806187 &      1578.304077 &             2.806187 &  1619.414551 &      2.705510 \\
 1899.369291 &  1594.143188 &      2.734031 &      1581.881104 &             2.734031 &  1619.858887 &      2.635603 \\
 1899.390125 &  1593.649170 &      2.682646 &      1582.187988 &             2.682646 &  1615.205566 &      2.580121 \\
 1899.410959 &  1593.457153 &      2.635216 &      1582.778320 &             2.635216 &  1616.738770 &      2.531829 \\
 \ldots &  \ldots & \ldots & \ldots & \ldots &  \ldots & \ldots \\
\enddata
\tablecomments{\obj\ light-curve data extracted from TESS sector 22 using DIA (corrected and uncorrected; see text for details) and SAP. Flux and flux error values of $-1$ indicate bad epochs that should be discarded in analysis. The full table is available in the online version.
\label{tab:1}}
\end{deluxetable*}


\subsection{Light-curve Modeling}


It has become increasingly popular to model AGN light curves in the time domain (as opposed to the frequency domain) with a Continuous Auto-Regressive Moving Average (CARMA) model \citep[e.g.,][]{Kelly_etal_2014,Simm_etal_2016,Caplar_etal_2017}. The CARMA models are a set of flexible Gaussian-process models to describe stochastic time series, with two parameters, $p$ and $q$, describing the orders of the auto-regression (AR) part and the moving average (MA) part. The DRW model is the lowest order CARMA model with $p=1$ and $q=0$, or a CAR(1) process. The major advantage of the CARMA model is that it can provide an accurate description of the stochastic light curve and the underlying PSD, while taking into account measurement uncertainties with robust Bayesian inference. The CARMA model is fit within the framework of a Gaussian process, which yields robust parameter and uncertainty estimation via sampling the posterior distribution. This procedure follows the standard Bayesian inference by incorporating the prior and the likelihood, and uses Markov Chain Monte Carlo techniques to sample the posterior. Importantly, since the modeling is performed in the time domain rather than in the frequency domain, the inferred PSD is more robust against windowing effects (such as aliasing and red noise leakage) in traditional power spectral analysis. 

However, it is important to note that a full CARMA model has the flexibility to introduce artificial curvatures in the PSD given the model prescriptions and the limited baseline of the light curve. It is difficult to assign physical meanings to the curvatures in the CARMA PSDs beyond the DRW model. In the case of the DRW model, however, the damping timescale $\tau_{\rm DRM}$ can be associated with the physical timescale over which the auto-correlation of the stochastic process is exponentially suppressed \citep[e.g.,][]{Kelly_etal_2009,Kozlowski_etal_2010,MacLeod_etal_2010}. Even though the nature of the damping process is still unclear, \cite{Kelly_etal_2009} suggest $\tau_{\rm DRM}$ may be related to the thermal timescale of an accretion disk under the assumption that the variability arises from thermal fluctuations in the accretion disk.

For these reasons, we focus our fiducial analysis on the DRW model, using the implementation in the public code \textsc{Javelin} \citep[][]{Zu_etal_2011}. But we also tested the full CARMA models using the public \textsc{carma\_pack} software \citep{Kelly_etal_2018}, and found that the best-fit CARMA model provides nearly identical results to the DRW model (see Figure~\ref{fig:carma} and \S\ref{sec:results}). 


\section{Results} \label{sec:results}

\begin{figure*}
\gridline{\fig{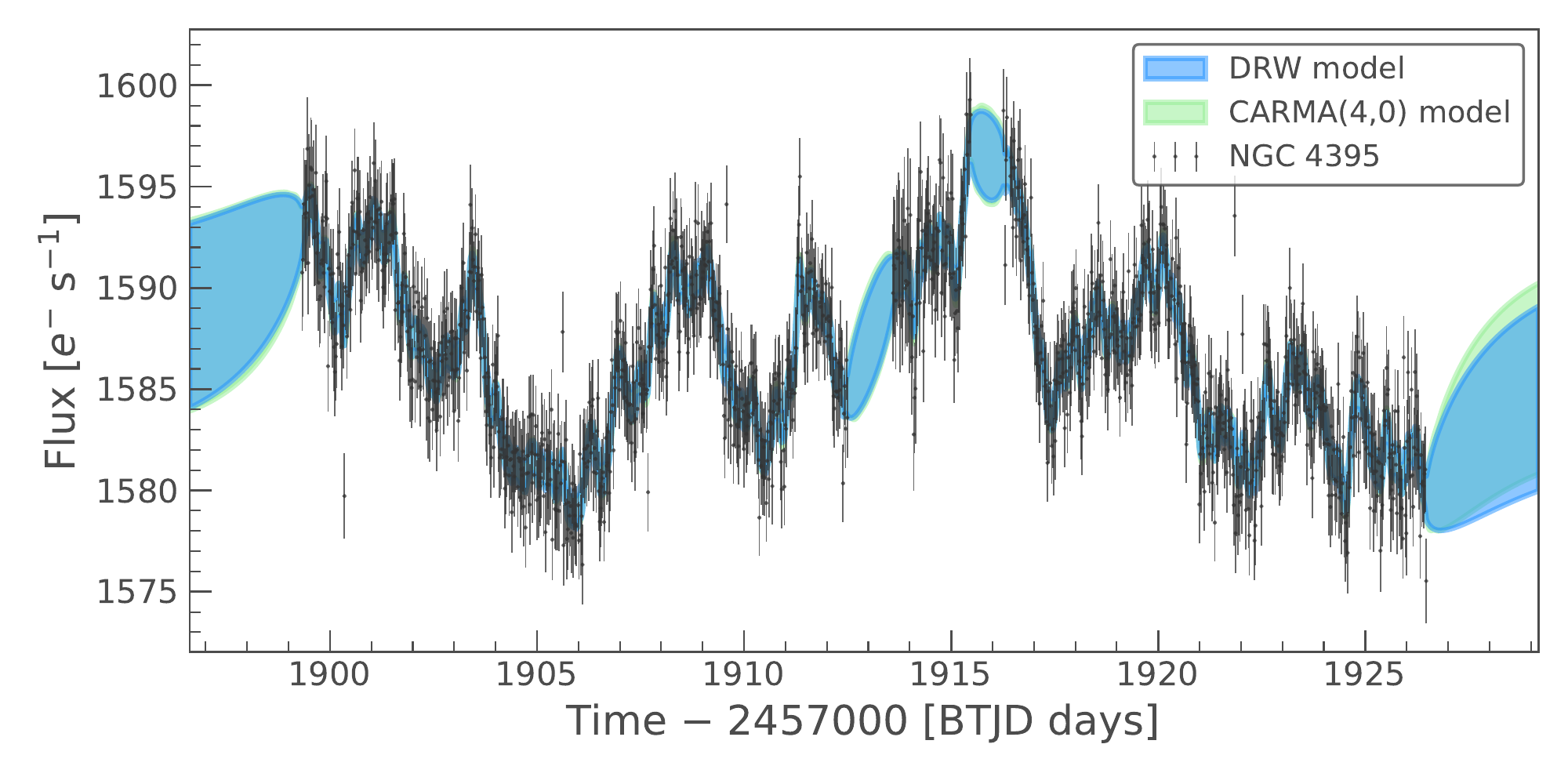}{0.95\textwidth}{(a)}
          }
\gridline{\fig{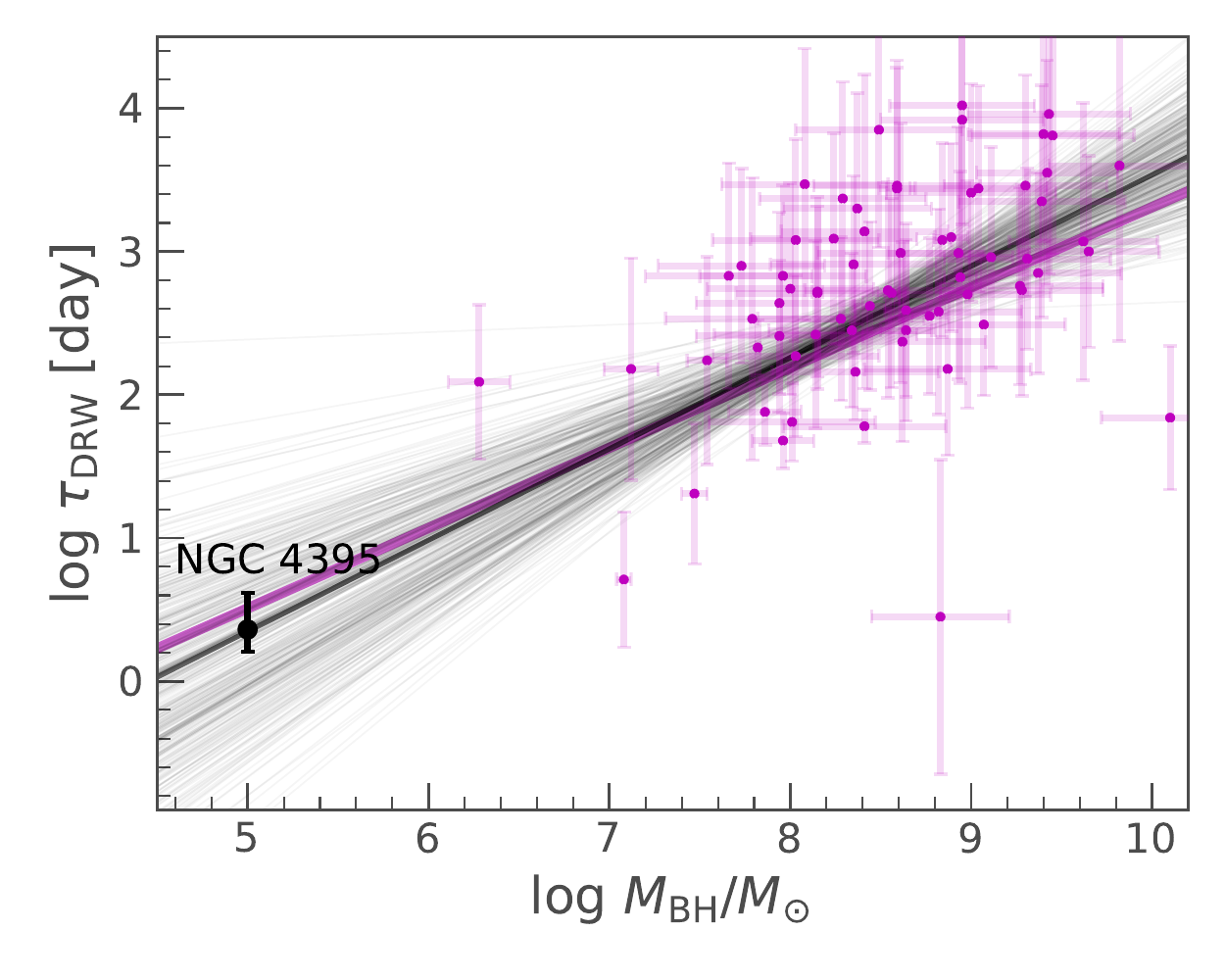}{0.49\textwidth}{(b)}
          \fig{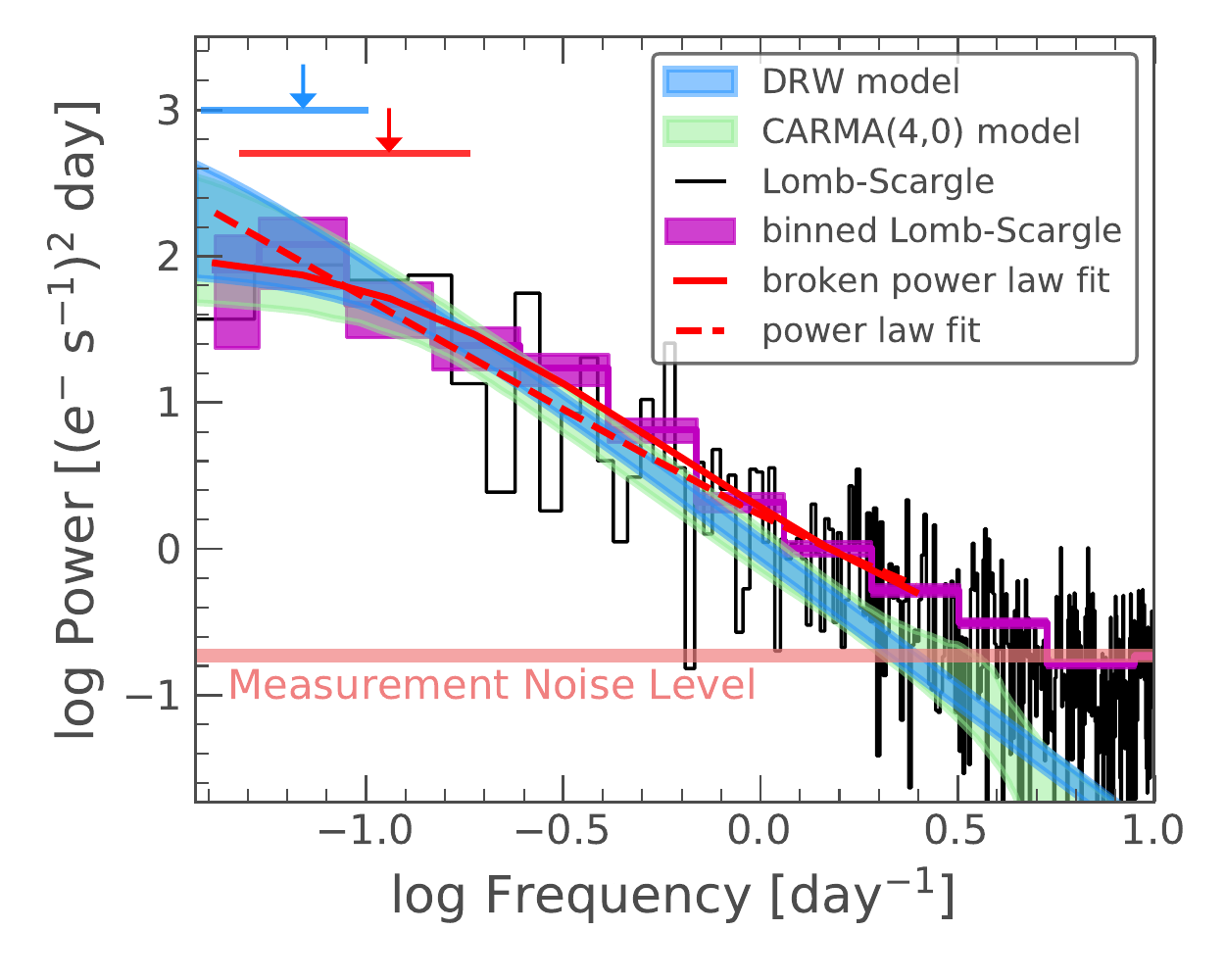}{0.49\textwidth}{(c)}
          }
\caption{\emph{Panel a)} Final, detrended DIA light curve of \obj\ with fitted DRW (CAR(1), green) and CARMA(4,0) (blue) models. The light shaded regions bracket the 1$\sigma$ error ellipse of the model. \emph{Panel b)} $\tau_{\rm DRW}$ vs. $M_{\rm{BH}}$. The AGN sample and best-fit relation from \citet{Kelly_etal_2009}, $\log(\tau_{\rm DRW}/{\rm day})=-2.29+0.56\log (M_{\rm BH}/M_\odot)$, are shown in magenta. The dark gray line is our re-fitting of the \citet{Kelly_etal_2009} data (excluding \obj) following the method of \citet{Kelly_2007} with 100 random draws from the posterior shown as gray lines. Our re-fitting is consistent with \citet{Kelly_etal_2009} given the stochasticity in the fitting. \obj\ lies well on the extrapolation of this relation. \emph{Panel c)} Power spectral density (PSD) for NGC 4395. The light shaded blue and green regions show the 95\% confidence range of the DRW and the best-fit full CARMA model (CARMA(4,0)) PSDs, which represent the intrinsic PSD and extend below the measurement noise level. The black histogram is the unbinned Lomb--Scargle periodogram, and the purple blocks are the logarithmically binned periodogram. The vertical extent of the purple blocks indicates the 1$\sigma$ uncertainties. The red dotted and dashed lines are the single power-law and broken power-law fits to the binned periodogram for frequencies $\log f<0.5\,{\rm day^{-1}}$. The break frequency from the DRW model (blue arrow with 1$\sigma$ uncertainty indicated by the blue segment) is consistent with the break frequency from the broken power-law fit to the binned periodogram (red arrow with 1$\sigma$ uncertainty indicated by the red segment). \label{fig:carma}}
\end{figure*}

\begin{figure*}
\fig{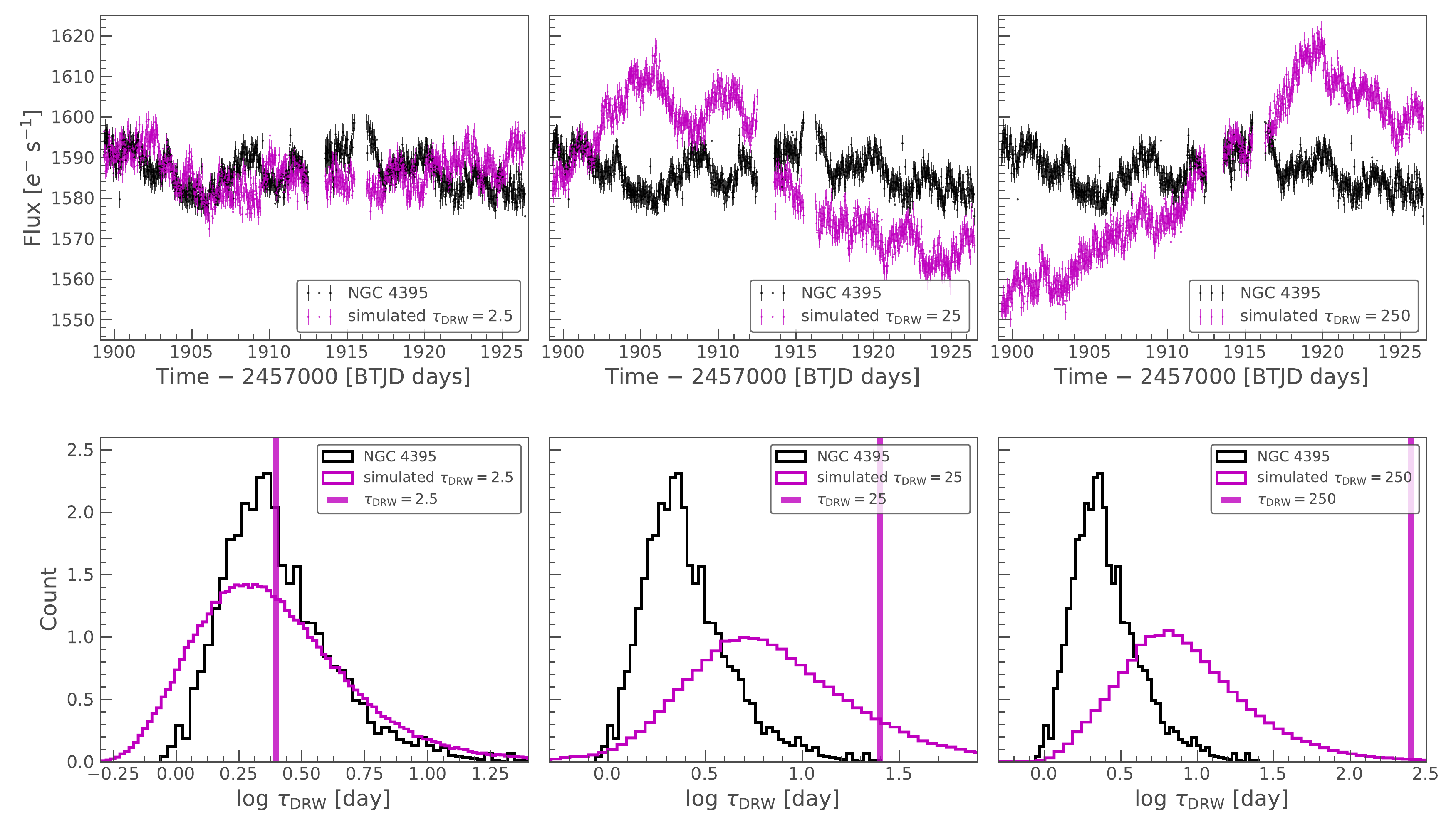}{0.99\textwidth}{}
\caption{\emph{Top row:} Final, detrended DIA light curve of \obj\ (black) and example simulated DRW light curves (magenta). While the variability amplitude on days timescales is similar in all three simulated cases of different input $\tau_{\rm DRW}$ values, cases with longer $\tau_{\rm DRW}$ values show notably more power on $\sim$week timescales. \emph{Bottom row:} Histograms of the $\tau_{\rm{DRW}}$ posterior distributions from the \textsc{Javelin} DRW fitting of \obj\ (black) and the simulated light curves (magenta) with areas normalized to 1. The true value of $\tau_{\rm{DRW}}$ (indicated by the purple vertical line) is accurately recovered for simulated light curves with $\tau_{\rm{DRW}}=2.5$ days. For cases where $\tau_{\rm DRW}$ is comparable to or longer than the duration of the light curve, the DRW fits with \textsc{Javelin} are unreliable and converge to $\sim 20-30\%$ of the light-curve duration regardless of the input $\tau_{\rm DRW}$ \citep[e.g.,][]{Kozlowski_2017}.\label{fig:sim}}
\end{figure*}


Figure~\ref{fig:carma} presents the main results of our light-curve analysis for \obj. The top panel displays the DIA light curve and the DRW model. The best-fit full CARMA model, CARMA(4,0), identified following the approach described in \citet{Kelly_etal_2014}, is also overplotted for comparison. The best-fit $\tau_{\rm DRW}$ from the DRW model is $2.3_{-0.7}^{+1.8}$\,days, corresponding to a transition frequency between a $f^{-2}$ PSD and a white noise PSD at $f_{\rm DRW}=1/(2\pi \tau_{\rm DRW})$. The bottom-left panel displays the correlation between $\tau_{\rm DRW}$ and $M_{\rm BH}$ measured in \citet{Kelly_etal_2009} for a sample of much more massive AGNs. \obj\ lies well on the extrapolation of this relation over three decades in BH mass. The bottom-right panel shows the optical PSD. The PSD from the DRW model is roughly consistent with the Lomb--Scargle periodogram \citep{Lomb_1976,Scargle_1982} directly computed from the light curve without accounting for light curve noise. 

In this work, we quantify any characteristic timescales in the variability using the DRW modeling performed in the time domain. Identifying a characteristic timescale (or frequency) in the frequency domain via periodogram analysis is more demanding because of windowing effects. Any frequency break in the periodogram near the lowest frequency enabled by the length of the light curve will likely be poorly constrained. Nevertheless, we fit the periodogram with simple power-law models to provide a sanity check on our DRW modeling results. 

To fit the periodogram we first bin it logarithmically, shown as the purple blocks in Figure~\ref{fig:carma}(c), where the vertical extent of the blocks indicates the 1$\sigma$ uncertainties of the power based on the number of points contributed to the bin. We then fit the binned periodogram both with a single power law, $P\propto f^{-\alpha}$, and with a broken power law, $P\propto 1/[(f/f_{\rm br})^\alpha + (f/f_{\rm br})^\beta]$, for frequencies below the noise regime ($\log f\,[{\rm day^{-1}}]<0.5$). Before the fit, we also subtract a constant noise power from the binned periodogram, which is required for more robust constraints on the high-frequency slope. 

The single power-law fit yields a slope of $\alpha=1.51\pm0.15$ with $\chi^2/{\rm dof}=13.96/7$, indicating a poor fit. The broken power-law fit yields a high-frequency slope $\alpha=1.88\pm0.15$, a low-frequency slope $\beta=0.00\pm 0.86$, and a break frequency $f_{\rm br}=0.114\pm0.066\,{\rm day}^{-1}$, with $\chi^2/{\rm dof}=3.28/5$. An F-test yields a significance at ${>}97$ percent in favor of the broken power-law model over the single power-law model. Visual inspection also favors the broken power-law fit to the periodogram. The broken power-law model constrains a high-frequency slope and a break frequency consistent with the DRW model. However, the constraint on the break frequency from the broken power-law fit is not great, reflecting the difficulty of finding a break frequency near the low-frequency end using periodogram analyses.  


The orbital frequency for the Innermost Stable Circular Orbit (ISCO) for \obj\ is $\log f[{\rm day^{-1}}]\approx 3.28$ (or $1/f=46$ seconds) at 6$R_g$ assuming a Schwarzschild BH with $M_{\rm BH}=10^5\,M_\odot$, which is deep in the frequency regime below the measurement noise level in the TESS PSD. For reference, the orbital frequency at 600$R_g$ is $\log f[{\rm day^{-1}}]\approx 0.28$, near the onset of our measurement noise floor. The optical PSD slope is also steeper than the X-ray PSD slope \citep[][]{Vaughan_etal_2005} over the small overlapping frequency range (around a few hours). This is somewhat expected given the different emission regions probed by optical and X-ray data. The X-ray emission is confined to a region of a few $R_g$ of the BH, while the optical variability probes a much larger region of the accretion disk. Strong correlations between optical and X-ray variability have been observed in AGN, including Seyfert galaxies \citep[e.g., ][]{Noda2016}, but no consensus has emerged as to the nature of and connection between the X-ray and optical variability \citep[e.g., ][]{Simm_etal_2016}.


The DRW model can result in unreliable $\tau_{\rm DRW}$ constraints if the true damping timescale is much longer than the duration of the light curve. \citet{Kozlowski_2017} has shown that reliable constraints on $\tau_{\rm DRW}$ can only be achieved when the duration of the light curve is at least ten times the damping timescale. For shorter light-curve lengths, the DRW fit will typically return unreliable $\tau_{\rm DRW}$ measurements that are $\sim 20-30\%$ of the duration of the light curve. Our measured $\tau_{\rm DRW}$ is less than 10\% of the duration of our TESS light curve, but it is still possible that the underlying process has a much longer damping timescale. We perform simulations to evaluate the robustness of our measured $\tau_{\rm DRW}$.

We generate mock light curves using the DRW model with the same cadence/duration as the TESS light curve. We test three cases of $\tau_{\rm DRW}=2.5, 25, 250$\,days, with the variability amplitude on days timescales matched to that of the TESS light curve. The different values of $\tau_{\rm DRW}$ then translates to different variability power on longer timescales ($\gtrsim 1$ week). The top three panels in Figure~\ref{fig:sim} show one random realization of the simulated light curves for the given $\tau_{\rm DRW}$ value. For these specific examples, the $\tau_{\rm DRW}=2.5$\,days case very much resembles the observed \obj\ light curve, while the other longer $\tau_{\rm DRW}$ cases show notably more power on longer timescales probed by the light curve. At face value, this already demonstrates that $\tau_{\rm DRW}=2.5$\,days is more consistent with the observed \obj\ light curve. However, since AGN light curves are stochastic, it is possible for the other two $\tau_{\rm DRW}$ cases to produce light curves that look similar to the real light curve when the duration is truncated. The bottom three panels show the posterior distribution of $\tau_{\rm DRW}$ from \textsc{Javelin} fits, stacked over 100 realizations of mock light curves for each input $\tau_{\rm DRW}$ case. We found that indeed the duration of TESS light curve is insufficient to recover $\tau_{\rm DRW}$ values that are longer than the observed duration. For these unconstrained DRW models, the measured $\tau_{\rm DRW}$ converges to $\sim 20-30\%$ of the observed light-curve duration, consistent with the findings in \citet{Kozlowski_2017}. 

Nevertheless, if the actual $\tau_{\rm DRW}$ were much longer than a few days, the most likely outcome from a DRW fit to the TESS light curve would be around $\sim 6-8$\,days, longer than the measured value of $2.3_{-0.7}^{+1.8}$\,days, and the posterior distribution of $\tau_{\rm DRW}$ would look markedly different from that measured from the light curve of \obj. Although we cannot rule out this possibility entirely, our results are more consistent with the idea that we are indeed measuring a very small $\tau_{\rm DRW}$ for \obj. This is further supported by the break frequency measured from the periodogram directly.

As a final test, we applied the same periodogram analysis to the mock DRW light curves simulated with an input $\tau_{\rm DRW}=2.5$\,days. In the vast majority of cases, the DRW modeling approach can robustly recover (e.g., within $\pm 1.5$\,days of) the input damping timescale (see Figure~\ref{fig:sim}). However, in most of these cases the periodogram analysis failed to recover the input damping timescale; in the remaining cases where the periodogram analysis revealed a break frequency consistent with the input $\tau_{\rm DRW}$, the constraint on the break frequency is at similar significance levels as that in the periodogram analysis on the real \obj\ light curve. This result strengthens our argument earlier that fitting in the time domain with rigorous Bayesian approach is more reliable to constrain the characteristic timescale (frequency) than simple periodogram analysis in Fourier space.   

\section{Discussion and Conclusions} \label{sec:discussion}


We have presented a month-long optical light curve of \obj\ from the TESS satellite. Variability is well detected from weeks to hours timescales. The high-quality light curve can be well fit by a DRW model, with a damping timescale $\tau_{\rm DRW}=2.3_{-0.7}^{+1.8}$\,days (1$\sigma$). The periodogram of the light curve can be better fit by a broken power law than a single power law, with a high-frequency slope $-1.88\pm0.15$, and a break frequency consistent with the measured $\tau_{\rm DRW}$ within $1\sigma$. The constraint on a damping timescale in the DRW modeling (and less so in the periodogram analysis) is made possible by the month-long duration of the TESS light curve. Shorter light curves will not be able to constrain this damping timescale well for \obj. It is still possible that the light curve is produced by a DRW process with a significantly larger $\tau_{\rm DRW}$ over the limited duration of the TESS light curve. Longer baselines of monitoring for \obj\ can help enhance or falsify our conclusions.

\obj\ also has well-measured X-ray variability PSDs \citep[e.g.,][]{Shih_etal_2003,Vaughan_etal_2005}, which probes the emission much closer to the BH than the optical emission from the accretion disk. The X-ray PSD reveals a break at a frequency of ${\sim}(0.5-3)\times 10^{-3}\,$Hz \citep{Vaughan_etal_2005}. The TESS optical PSD cannot probe this high-frequency break shown in the X-ray PSD ($\log f\,[{\rm day^{-1}}]\approx 1.9$) due to background noise and insufficient sampling below hourly timescales. For the same reason, we do not observe a steepening at the highest frequencies in the optical PSD, as observed for more massive ($M_{\rm BH}\gtrsim 10^7\,M_\odot$) AGNs on timescales below $\sim 2$ days \citep[e.g.,][]{Mushotzky_etal_2011,Kasliwal_etal_2015}. Given the 2-3 orders of magnitude lower BH mass in \obj, the expected steepening in the optical PSD would occur well within the regime dominated by measurement noise.


Nevertheless, the month-long TESS optical light curve of \obj\ tentatively revealed, for the first time, a damping timescale on much longer timescales (or much lower frequencies) than in the X-rays. The measured damping timescale $\tau_{\rm DRW}=2.3_{-0.7}^{+1.8}$\,days from the DRW model places \obj\ on the extrapolation of the empirical relation between BH mass and $\tau_{\rm DRW}$ measured for much more massive SMBHs \citep[][]{Kelly_etal_2009}. This opens the possibility of using optical variability of AGN to constrain the SMBH mass, if this $\tau_{\rm DRW}-M_{\rm BH}$ correlation holds over the full range of SMBH masses ($\sim 10^4-10^{10}\,M_\odot$). Our results resonate with the findings in \citet{Smith_etal_2018}, where they measured the break in the optical PSD for six AGNs and found a correlation between the characteristic timescale in the PSD and the mass of the BH. More general applications of characteristic timescales in optical variability to various accreting systems are also promising \citep[e.g.,][]{Scaringi_etal_2015}. 

\obj\ is used in this work as a proof of concept to demonstrate the feasibility of our imaging analysis of TESS data and applications to extragalactic science. In future work, we will explore the full potential of TESS light curves in discovering nuclear variability in bright galaxies and measuring characteristic timescales in the light curve to correlate with SMBH mass.  


\acknowledgments

C.J.B. acknowledges the Illinois Graduate Survey Science Fellowship for support. Y.S. acknowledges support from an Alfred P. Sloan Research Fellowship and NSF grant AST-1715579. C.A.F.G. was supported by the NSF through grant AST-1715216 and CAREER award AST-1652522; by NASA through grant 17-ATP17-0067; and by a Cottrell Scholar Award from the Research Corporation for Science Advancement. Financial support for this publication results partially from a Scialog program sponsored by Research Corporation for Science Advancement (Y.S., S.S., C.A.F.G.). We thank Charles Gammie and Jennifer Li for useful discussions. We thank the anonymous referee for useful comments which improved this work. We thank Ann Wehrle for helpful comments which improved Figure~\ref{fig:images}.

This paper includes data collected by the TESS mission. Funding for the TESS mission is provided by the NASA Explorer Program. 

%

\vspace{5mm}
\facilities{TESS}


\software{astropy \citep{Astropy_2018},  
          \textsc{hotpants} \citep{Becker_2015},
          lightkurve \citep{Lightkurve_Collaboratioin_2018}, 
          astroquery \citep{Ginsburg_etal_2019},
          carma\_pack \citep{Kelly_etal_2014,Kelly_etal_2018},
          Javelin \citep{Zu_etal_2011},
          reproject \citep{robitaille_thomas_2018},
          matplotlib \citep{Hunter_2007}
          }



\appendix

\section{Difference Imaging Method} \label{sec:kernel}

Our DIA method utilizes \textsc{hotpants}, which follows the algorithm described in \citet{Alard_Lupton_1998} and \citet{Alard_2000}. This code works by minimizing the equation,

\begin{equation}
    \sum_i \left( \left[T \otimes K \right](x_i,y_i) - I(x_i,y_i) \right)^2
\end{equation}

where $T$ is the template frame, $K$ is the PSF kernel, $I$ is the science frame, and $\otimes$ denotes convolution. We assume the kernel $K$ can be decomposed into Gaussian basis functions which are allowed to vary on differing spatial orders. This takes the form,

\begin{equation}
    K(u,v)=\sum_n a_nK_n(u,v)
\end{equation}
where,
\begin{equation}
    K_n(u,v)=e^{-\left(u^2+v^2\right)/2\sigma_k^2}u^iv^j\ ,
\end{equation}
and $n=(i,j,k)$.

We use,
\[  \sigma_k = \left\{
\begin{array}{ll}
      0.7 & \text{if}\ i+j\leq8 \\
      1.5 & \text{if}\ i+j\leq6 \\
      3.0 & \text{if}\ i+j\leq4 \\
      8.0 & \text{if}\ i+j\leq2 \\
\end{array} 
\right. \]

The spatial order is confined to the size of kernel. We adopt a kernel size of $11\times11$ pixels which is well-matched to the PSF size of TESS. The assumption is that sources used to build the kernel are isolated and non-variable to normalize the flux to the template frame. This removes most long-term systematic trends in the light curves, although some residual trends remain due to imperfect background subtraction.

To evaluate the quality of the image subtraction in each frame, we also plot the normalized residuals of each difference frame and inspect them visually. For sector 22 frames of \obj\, frames with biased residuals lie in the first days of each orbit when the spacecraft is in ``thermal recovery'' from an Earth pointing and frames suffering from rolling-band noise.

\section{Residual Detrending} \label{sec:detrending}

\begin{figure*}
\fig{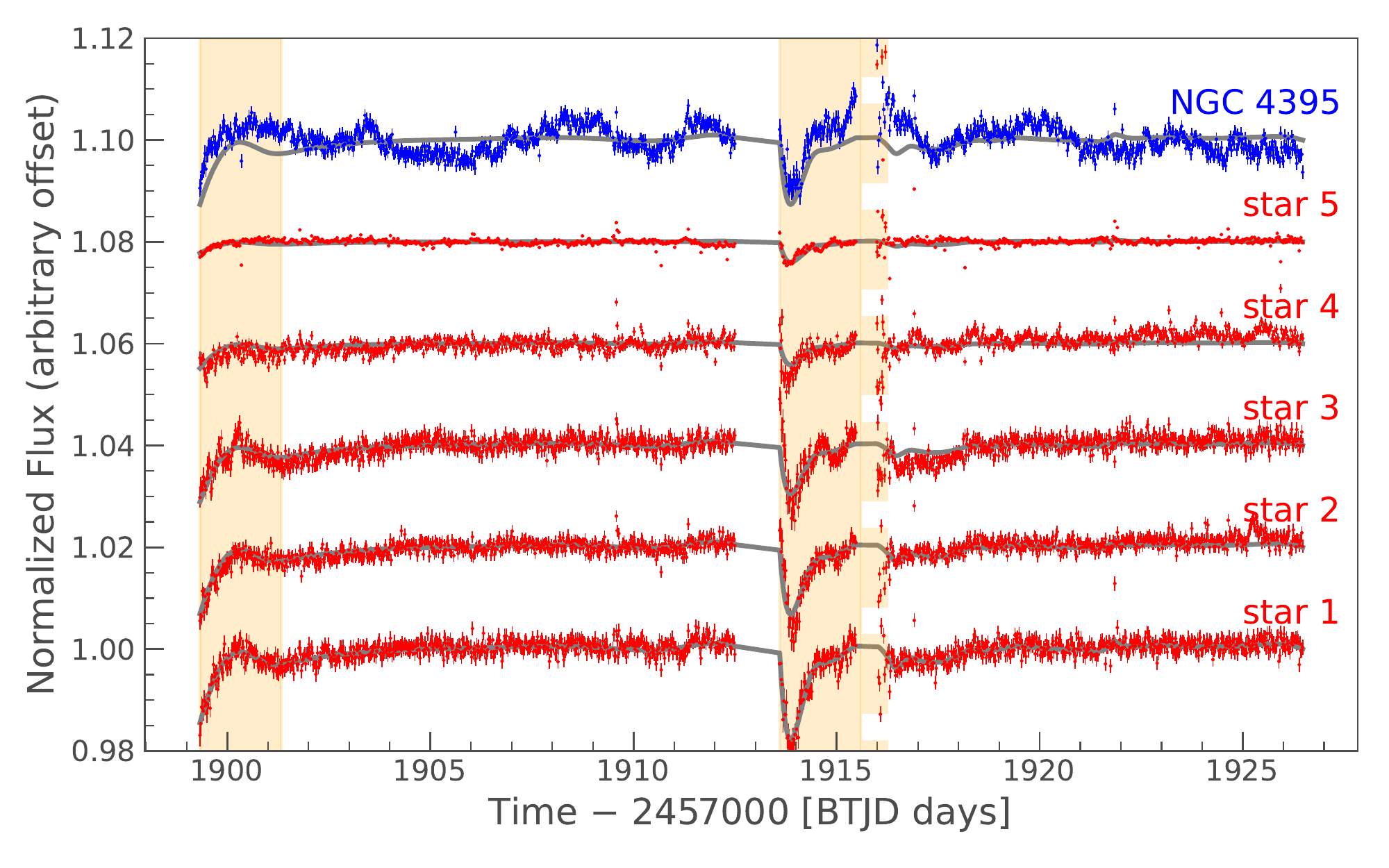}{0.95\textwidth}{}
\caption{Uncorrected DIA light curves of the target (blue, top row) and nearby field stars (red) shown on a normalized flux scale. Best-fit spline detrending models are shown as solid gray lines. The solid orange shaded areas show the first day of each orbit: the thermal recovery period when residual systematic trends tend to be large. The frames in the dashed orange shaded areas are affected by rolling-band noise. The small gaps in the photometry here are bad frames flagged automatically by the TESS pipeline. Rolling-band noise is difficult to correct, so these frames were simply discarded in the corrected light curve. \label{fig:lc_uncorrected}}
\end{figure*}

To model residual trends in the light curves, we select star 1 as our ``calibration'' reference star. We use star 1 as the calibration star because it is impacted most strongly by the common systematic dips displayed in the light curves. This is likely a function of the fractional background flux included in the chosen aperture relative to the star's flux. We fit two smoothing splines, $\mathbf{s_1}$ and $\mathbf{s_2}$, to the star 1 light curve in first and second orbit. Each spline is a piecewise function of $n-1$ 3rd-order polynomials fit to star 1 with the condition that they connect at each knot $n$. We found $n=450$ for orbit 1 and $n=550$ for orbit 2 was able to fit the dips near the beginning of each orbit without overfitting to the scatter in the photometry.

We then use the splines derived from star 1 to construct a generalized model to fit to each light curve. The systematic trends in each light curve have varying strength depending on the flux of the sources and choice of aperture. Therefore, we wish to preserve the shape of the trends in each orbit while allowing for re-scaling/stretching. We minimize the $\chi^2$ within the first three days of each orbit using the residual model,
\begin{equation}
\hat{\mathbf{y}}_{\rm{res}} = a_1 \mathbf{s_1}(\mathbf{t_1}) + a_2 \mathbf{s_2}(\mathbf{t_2}) + b
\end{equation}

where $\mathbf{s_1}(\mathbf{t_1})$ and $\mathbf{s_2}(\mathbf{t_2})$ are the splines in the first and second orbits derived from star 1 interpolated at times $\mathbf{t_1}$ and $\mathbf{t_2}$. We fit the re-scaling coefficients $a_1$ and $a_2$ to each orbit in the light curve and a median flux offset $b$. By fixing the shape of the model to the trends in the calibration star light curve, our method has the advantage of preserving intrinsic trends in the target light curve. The maximum likelihood estimation is only done inside the orange windows shown in Figure~\ref{fig:lc_corrected} and Figure~\ref{fig:lc_uncorrected} (corresponding to the first day of each orbit) to fit the strength of the residual dips near the beginning of each orbit. We also remove 3$\sigma$ outliers from each light curve at this stage. The models fitted to each uncorrected light curve are shown in Figure~\ref{fig:lc_uncorrected}, and the resulting corrected light curves are shown in Figure~\ref{fig:lc_corrected}.

\bibliography{refs,refs_YS}{}
\bibliographystyle{aasjournal}



\end{document}